\documentstyle[aps,12pt]{revtex}

\begin{document}
\title{Comment on ''Low to High Spin-state Transition Induced by charge
ordering in Antiferromagnetic YBaCo$_{2}$O$_{5}$'' }
\author{Z. Ropka}
\address{Center for Solid State Physics, S$^{nt}$ Filip 5, 31-150 Krak\'{o}w.%
}
\author{R.J. Radwanski}
\address{Center for Solid State Physics, S$^{nt}$ Filip 5, 31-150 Krak\'{o}%
w, \\
Institute of Physics, Pedagogical University, 30-084 Krak\'{o}w, POLAND.\\
email: sfradwan@cyf-kr.edu.pl}
\maketitle

\begin{abstract}
In the commented Letter Vogt et al. (Phys.Rev.Lett. 84 (2000) 2969) claim
the existence of a spin-state transition in YBaCo$_{2}$O$_{5}$. We think
that this claim about the change from the low spin to high spin state upon
cooling in YBaCo$_{2}$O$_{5}$ is errorneous.


PRL\ submit: 31-07-2000
\end{abstract}

\date{}

In Ref. 1 Vogt et al. claim the existence of a spin-state transition in YBaCo%
$_{2}$O$_{5}$. The spin state changes from low spin to high spin upon
cooling. We think that this claim is errorneous. In Ref. 1 this claim has
been derived on basis of

A) the comparison of the neutron-diffraction derived moment of 2.10 $\mu
_{B} $ per average cobalt ion at 300 K with the neutron-diffraction derived
moment of 2.7 $\mu _{B}$ and 4.2 $\mu _{B}$ at 25 K. We will denote the
magnetic moment derived by neutron-diffraction experiments by m$_{n}.$

B) a theoretical analysis of the value of 2.10 $\mu _{B}$ found at 300 K as
corresponding ''to the spin-only value expected for a mixed valence compound
containing equal concentrations of low spin (LS)\ Co$^{2+}$ ($^{6}$t$_{2g}$ $%
^{1}$e$_{g}$, S=1/2, $\mu _{so}$ = 1.73 $\mu _{B}$) and intermediate spin
(IS) Co$^{3+}$ ($^{5}$t$_{2g}$ $^{1}$e$_{g}$, S=1, $\mu _{so}$ = 2.85 $\mu
_{B}$)''. Values of 2.7 and 4.2 $\mu _{B}$ found at 25 K have been
interpreted in Ref. 1 as related with ''IS Co$^{3+}$ ($^{5}$t$_{2g}$ $^{1}$e$%
_{g}$, S=1, $\mu _{so}$ = 2.85 $\mu _{B}$) and HS (high spin) Co$^{2+}$ ($%
^{5}$t$_{2g}$ $^{2}$e$_{g}$, S=3/2, $\mu _{so}$ = 3.87 $\mu _{B}$)'' or to
an alternative composition of ''HS Co$^{3+}$ ($^{4}$t$_{2g}$ $^{2}$e$_{g}$,
S=2, $\mu _{so}$ = 4.89 $\mu _{B}$) and HS Co$^{2+}$ ($^{5}$t$_{2g}$ $^{2}$e$%
_{g}$, S=3/2, $\mu _{so}$ = 3.87 $\mu _{B}$)''.

According to us the behaviour described in point A is fully conventional for
the ordered magnetic moment as is the case of YBaCo$_2$O$_5$ with T$_N$ of
330 K - the ordered moment always increases with the decreasing temperature
being zero above T$_N$ [see, e.g. Ref. 2, ch.15 or Ref. 3 ch. 31]. Thus the
claim about the change of the spin state from the low spin to high spin upon
cooling is a misunderstanding. In the recalled by authors of Ref. 1 the case
of LaCoO$_3$ there is a transition from a high to low-spin state (via an
intermediate state) upon cooling [4,5], i.e. the moment at lower temperature
is lower than at high temperatures what makes LaCoO$_3$ the intriguing case.

The theoretical analysis mentioned as point B is even more errorneous. The
authors of the commented paper [1] errornously compare the value of the
ordered moment found in the neutron diffraction experiments m$_n$ with the
theoretical values for the effective moment observed in the paramagnetic
state. The former is proportional to the z component of the quantum number S
whereas the latter is related to the $\surd $S(S+1). For the calculation of
the magnetic moment we have to take into account g$_s$=2. It means that for
the LS Co$^{2+}$ ion described by S=1/2 one should expect the ordered moment
of 1 $\mu _B$ (m$_n$) whereas the effective moment p$_{eff}$ of 1.73 $\mu _B$
would be observed in the paramagnetic state in the temperature dependence of
the paramagnetic susceptibility. Similarily for S=1 we have m$_n$=2 $\mu _B$
but p$_{eff\text{ }}$= 2.83, for S=3/2 we have m$_n$=3 $\mu _B$ but p$_{eff}$
= 3.87, for S=2 we have m$_n$= 4 $\mu _B$ but p$_{eff}$ = 4.89 $\mu _B$
[Ref. 2, ch. 14 Table 2, Ref. 3 ch. 31 Table 31.4]. The value given in Ref.
1 for S=1 is slightly wrong (2.85 instead of 2.83 $\mu _B$). In such the
situation, the explanation for the value of 2.10 $\mu _B$ made in Ref. 1 is
not valid any more. As well as for others.

In the above discussion we do not touch at all the problem of the orbital
moment and the applicability of the spin-only theory to real 3d-ion systems.
This orbital moment has to be taken into account in order to explain the
ordered-moment value of 4.2 $\mu _{B}$ as the spin-only theory yields the
maximal value of 4 $\mu _{B}$ for Co$^{3+}$ and 3 $\mu _{B}$ for the Co$%
^{2+} $ ions. The moment of the needed value has been found recently for the
atomic-like 3d$^{6}$ systems, relevant to the Co$^{3+}$ and Fe$^{2+}$ ions,
taking into account the spin-orbit coupling that enables revealing of the
orbital magnetic moment [6-8]. Moreover, in Ref. 6 the very large influence
of the local symmetry on the local moment has been discussed showing that
not always the largest z-component of the angular momentum is realized as
the ground state what can lead to the non-magnetic ground state of the Co$%
^{3+}$ ion in the high-spin state S=2 like it is in LaCoO$_{3}.$

In conclusion, the claim of Ref. 1 about the change from the low spin to
high spin state upon cooling in YBaCo$_2$O$_5$ is errorneous.


\begin{references}
\bibitem{1} T.Vogt et al. Phys.Rev.Lett. 84, 2969 (2000).

\bibitem{2} C.Kittel, in: Introduction to Solid State Physics (1996), ch.14
and 15.

\bibitem{3} N.W.Aschroft and N.D.Mermin, in: Solid State Physics (1976) ch.
31.

\bibitem{4} M.A.Korotin et al. Phys.Rev. B 54, 5309 (1996).

\bibitem{5} P.Ravindran et al. Phys.Rev. B 60, 16423 (1999).

\bibitem{6} R.J.Radwanski and Z.Ropka, Solid State Commun. 112 (1999) 621.

\bibitem{7} R.J.Radwanski and Z.Ropka, Relativistic effects in the
electronic structure for the 3d paramagnetic ions,
http://xxx.lanl.gov/abs/cond-mat/9907140.

\bibitem{8} Z.Ropka, R.Michalski and R.J.Radwanski, Electronic and magnetic
properties of FeBr$_{2}$, http://xxx.lanl.gov/abs/cond-mat/0005502.
\end{references}
\end{document}